# Accelerated Evaluation of Automated Vehicles in Car-Following Maneuvers

Ding Zhao, Xianan Huang, Huei Peng, Henry Lam, David J. LeBlanc

***Abstract*—The safety of Automated Vehicles (AVs) must be assured before their release and deployment. The current approach to evaluation relies primarily on (i) testing AVs on public roads or (ii) track testing with scenarios defined in a test matrix. These two methods have completely opposing drawbacks: the former, while offering realistic scenarios, takes too much time to execute; the latter, though it can be completed in a short amount of time, has no clear correlation to safety benefits in the real world. To avoid the aforementioned problems, we propose Accelerated Evaluation, focusing on the car-following scenario. The stochastic human-controlled vehicle (HV) motions are modeled based on 1.3 million miles of naturalistic driving data collected by the University of Michigan Safety Pilot Model Deployment Program. The statistics of the HV behaviors are then modified to generate more intense interactions between HVs and AVs to accelerate the evaluation procedure. The Importance Sampling theory was used to ensure that the safety benefits of AVs are accurately assessed under accelerated tests. Crash, injury and conflict rates for a simulated AV are simulated to demonstrate the proposed approach. Results show that test duration is reduced by a factor of 300 to 100,000 compared with the non-accelerated (naturalistic) evaluation. In other words, the proposed techniques have great potential for accelerating the AV evaluation process.**

## I. Introduction

Automated vehicles (AVs) are being developed to shift the human driving function to robotic driving, and even possibly all the way to driverless automation [1]. Before their release and deployment, however, AVs must be thoroughly evaluated to ensure their safety.

Two fundamental questions arise regarding AV evaluation:

i) How to choose test scenarios
ii) How to interpret the test results to understand the real world performance of the AV.

We categorize the existing approaches to AV evaluation into four groups. Frequently, vehicle test protocols and test scenarios are predefined in the form of a "test matrix.". The scenarios are usually selected based on past crash data and expert knowledge. Many studies [2]–[10] have been undertaken to develop test procedures using this approach. Applying the test matrix method to evaluate low-level AVs is straightforward, and in the near future might continue to be the selected approach. However, a problem arises with driverless AVs because the vehicle is intelligent and can be calibrated to excel in the predefined tests, while performance in other test scenarios receives less attention [11].

A popular alternative to testing high automation level AVs is the Naturalistic-Field Operational Test (N-FOT), in which a number of equipped vehicles are tested under naturalistic driving conditions over an extended period of time [12]. In an N-FOT, the test scenarios are unpredicted, and the test results directly reflect real world performance. Several N-FOT projects [13]–[20] have been conducted in the U.S. and Europe. However, an obvious problem of the N-FOT approach is the time required. In the U.S., one needs to drive, on average, 530 thousand miles to experience a police-reported crash and nearly 100 million miles for a fatal crash, which is longer than the distance from the Earth to the Sun (93 million miles). In other words, a significant portion of N-FOT is "boring" and not effective testing. Because of the low exposure rate, N-FOT is unlikely to be used in government-approved tests.

Some researchers used Monte Carlo simulations [21]–[23] to emulate N-FOT tests by building stochastic models. Computer simulations can reduce the time and cost compared with field tests. However, low exposure to safety critical scenarios is still an issue. Using the original probability density function, the crude Monte Carlo method cannot accelerate simulations or tests.

Focusing on the safety-critical scenarios, the Worst-Case Scenario Evaluation (WCSE) methodology was studied in [24]–[26] with model-based optimization techniques. While the WCSE method targets the weakness of a vehicle control system to generate worst-case disturbances, it does not consider the probability of occurrence of the worst-case scenarios. Therefore, the worst case evaluation results do not provide information about risks in the real world. The four types of AV evaluation approaches are summarized in TABLE I. A more comprehensive analysis can be found in [27].

In this paper, we propose the Accelerated Evaluation (AE) approach. It starts from the Monte Carlo method but uses the Importance Sampling concept to result in much faster stochastic simulations. The most import benefit of the proposed

Ding Zhao and David J. LeBlanc are with the University of Michigan Transportation Research Institute, Ann Arbor, MI 481099 USA (e-mail: zhaoding@umich.edu, leblanc@umich.edu).

Xianan Huang and Huei Peng are with the Mechanical Engineering Department, University of Michigan, Ann Arbor, MI 48109 USA (e-mail:, xnhuang@umich.edu and hpeng@umich.edu).

Henry Lam is with the Department of Industrial and Operations Engineering, University of Michigan, Ann Arbor, MI 48109 USA (e-mail: khlam@umich.edu).

TABLE I
SUMMARY OF AV EVALUATION APPROACHES

| Approach | Basis | Examples | Advantages | Limitations |
|---|---|---|---|---|
| Test Matrix | Crash data<br>Expert knowledge | CAMP [2], HASTE [3], AIDE [4], TRACE [5], APROSYS [6], ASSESS [7], Euro-NCAP[8], Volpe [9] [10] | Efficient repeatable | Test scenarios are fixed and predefined.<br>The failure modes of AVs might not be reflected in the existing crash scenarios |
| Naturalistic Field Operational Tests | Public road tests | 100 Car Naturalistic Driving Study[28], ACAS [29], RDCW [13], SeMiFOT [15], IVBSS [16], SPMD [30], Google driverless car [31] | Real-world test | Inefficient<br>Expensive and time-consuming |
| Monte Carlo Simulation | N-FOT driving data | Yang *et al.* [21], Lee [22], Woodrooffe *et al.* [23] | Stochastic | Does not reduce non-safety critical events |
| Worst-Case Scenario Evaluation | Vehicle dynamics and control | Ma *et al.* [24], Ungoren *et al.* [25] and Kou [26] | Worst-Case scenarios | Does not relate to real-world driving conditions |

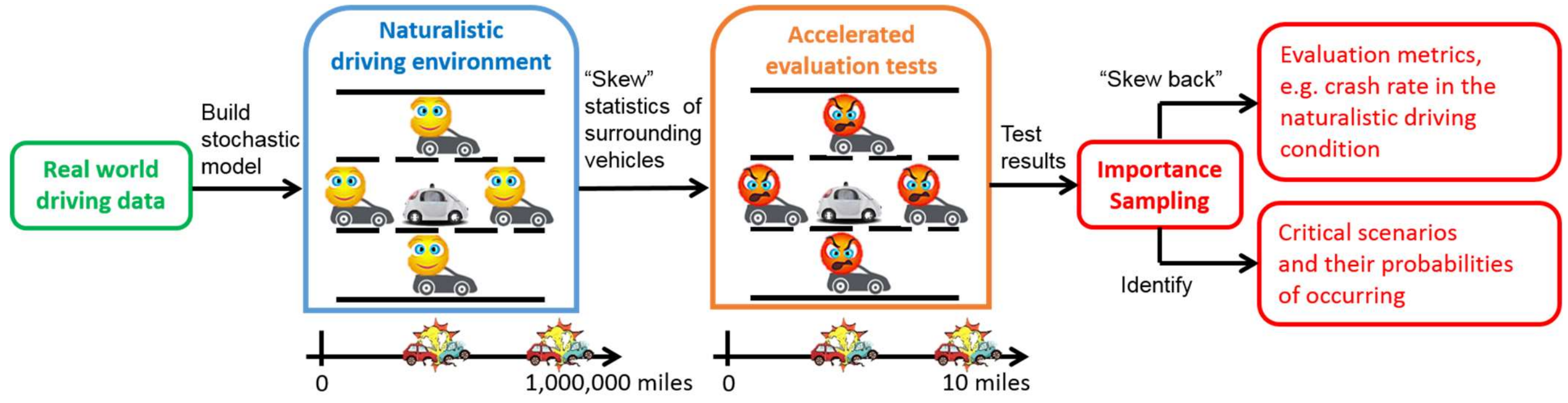

Fig. 1. Procedure of the Accelerated Evaluation method

method is that the real-world safety benefit of the AV can be assessed, while achieving an acceleration rate of hundreds to thousands of times greater compared to the original, non-accelerated, or so-called Crude Monte Carlo method.

We divide the AE method into four steps. As shown in Fig. 1, first, the behaviors of the "primary other vehicle" are modeled as the major disturbance to the operation of the AVs based on large-scale naturalistic driving data. Second, we skew the disturbance statistics to emphasize the safety-critical scenarios in daily driving. Third, the "accelerated tests" with modified statistics are conducted. Finally, we use Importance Sampling (IS) to "skew back" the results to understand real-world safety benefits (e.g., crash rate in the naturalistic driving conditions), and critical scenarios that may cause crashes and their probabilities of occurring.

In our previous work [32], [33], the AE approach is applied to study AV behavior in lane change scenarios. In this paper, we extend the AE concept to the car-following scenarios, which is a more challenging problem, as in the lane change scenarios, the randomness of the motion of the lane-changing vehicle is modeled as a set of distributions but sampled only once at the lane change moment. In the car-following scenario, the motion of the lead vehicle needs to be sampled from the stochastics for multiple steps. The statistics of the lead vehicle are state-dependent and thus change at each sampling step. We call this type of stochastic process "dynamic sampling". In [34], we proposed an empirical approach to evaluating AVs in the car-following scenario. This approach can generate critical scenarios and is useful for comparing between AV designs, but not yet able to rigorously estimate the crash rate of a new AV under naturalistic conditions and evaluate its safety benefit. To evaluate the AV's reaction to its leading human-controlled vehicle (HV), we develop a new Accelerated Evaluation approach by considering the correlations between each sample of the HV maneuvers. Three types of events—crash, injury, and conflict—are analyzed to demonstrate the effectiveness and robustness of this approach.

## II. MODEL OF THE CAR-FOLLOWING SCENARIOS

Car-following is one of the most fundamental driving tasks. In the Test Matrix method, AVs are tested with predefined lead vehicle motions including static, constant speed driving, and constant deceleration braking [8]. In this research, we model the lead HV as a stochastic model and test AV in a stochastic, dynamic environment.

### *A. Extraction of Naturalistic Car-following Events*

The data used to model car-following behaviors is from the Safety Pilot Model Deployment (SPMD) database [30]. The SPMD program aims to demonstrate connected vehicle technologies in a real-world environment. It recorded naturalistic driving of 2,842 equipped vehicles in Ann Arbor, Michigan for over two years. As of April 2016, 34.9 million miles were logged, making SPMD one of the largest public N-FOT databases ever.

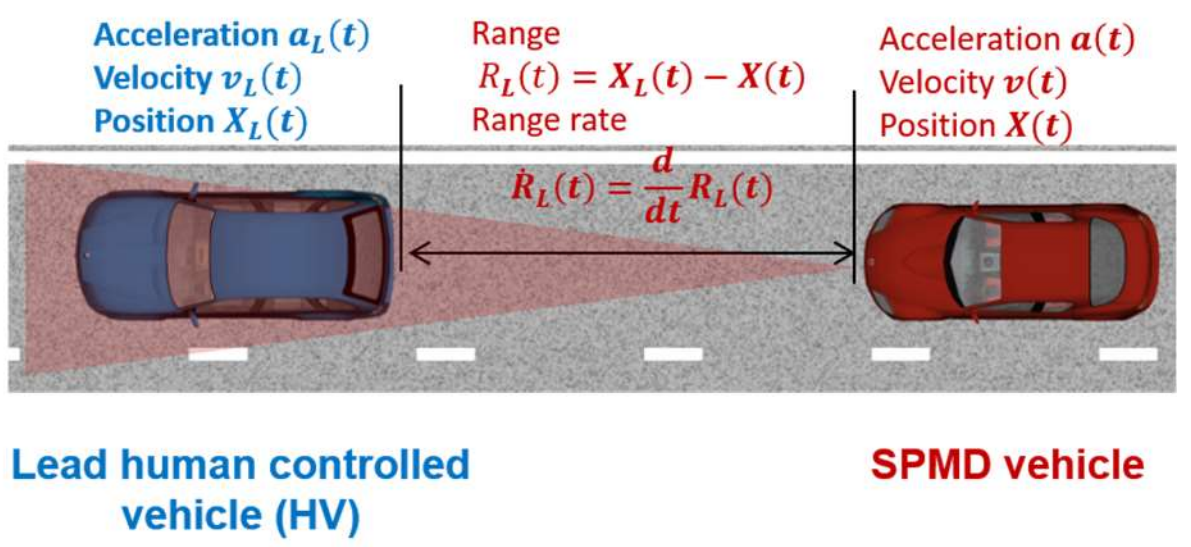

Fig. 2. Car-following scenarios captured by SPMD vehicles to measure the lead vehicle motion

The variables for a car-following scenario is defined in Fig. 2. In the SPMD program, 98 sedans are equipped with Data Acquisition System and MobilEye® [35], which provides: (a) relative position to the lead vehicle (range), and (b) lane tracking measures about the lane delineation both from the painted boundary lines and the road edge. The error of range measurement is around 10 % at 90 m and 5 % at 45 m [36].

To ensure consistency of the dataset employed, we apply the following criteria:

- $R_L(t) \in (0.1\text{ m}, 90\text{ m})$
- Longitude $\in (-88.2°,\ -82.0°)$
- Latitude $\in (41.0°,\ 44.5°)$
- No cut-in vehicles between HV and SPMD vehicle
- No lane changes by either vehicle
- Duration of car-following $> 50$ s

where $v_L$ and $v$ are the velocities of the lead HV and the SPMD vehicle; $R_L$ is the range, that is, the distance between the rear end of the HV and the front end of the SPMD vehicle. 163,332 car-following events were detected. Fig. 3 shows the locations of the identified car-following scenarios.

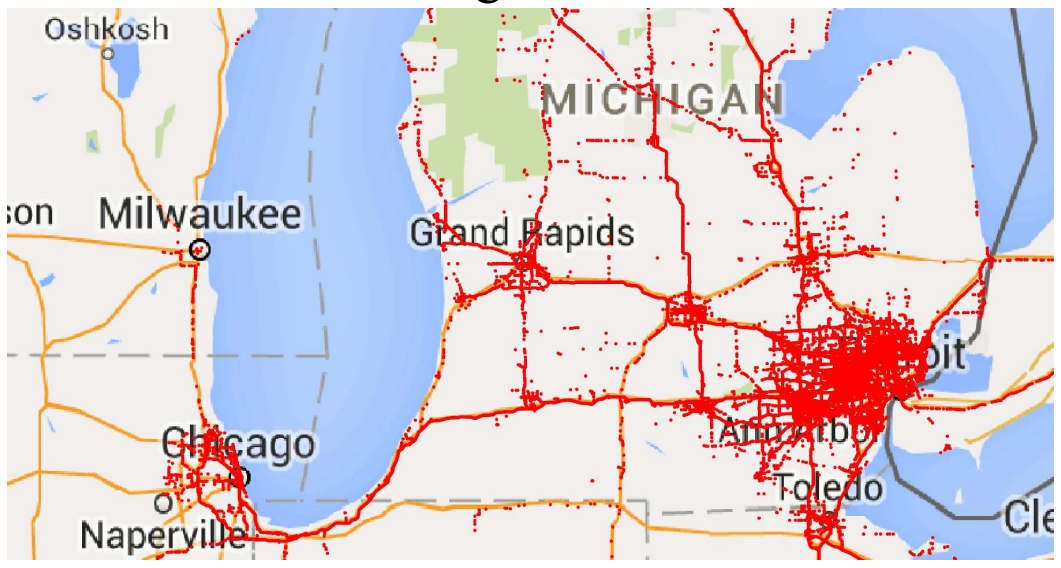

Fig. 3. Recorded car-following events from the SPMD database

## B. *Lead Human Controlled Vehicle Model*

We use the data captured in the SPMD database in the car-following scenarios to model the lead vehicle motion. In general, there are two types of driver models. The majority of driver models aim to predict driver behaviors deterministically, as in the models proposed in [37]–[39]. The other type of models captures the stochastic behavior of human drivers. The former type driver model is usually used in a real-time system such as the Advanced Driver Assistance Systems (ADAS), while the latter is more suitable for estimating the benefits in AV evaluation, for example, predicting how many crashes will occur for a new AV. The second type of AV model is the one adopted in this paper. The lead AV is modeled as a stochastic model similar to the one used in [40] and [34] to simulate the randomness of the lead vehicle. The vehicle acceleration in the next time step is predicted based on current step acceleration and velocity.

$$\begin{aligned} a_L(k+1) &= h_0 + h_1 a_L(k) + h_2 v_L(k) + u_h \\ &= [1, a_L(k), v_L(k)]\boldsymbol{h} + u_h \end{aligned} \quad (1)$$

where the driver model parameter vector $\boldsymbol{h} = [h_0, h_1, h_2]^T$ and $u_h \sim \mathcal{N}(0, \sigma_u^2)$ , representing the randomness in human behaviors.

The MobilEye®, Inertial Measurement Unit (IMU) and CAN bus on the SPMD vehicles provide 10 Hz measurement of $v$, $R_L$, acceleration $a$, and range rate

$$\dot{R}_L(k) = v_L(k) - v(k) \quad (2)$$

The lead HV velocity can be calculated from

$$v_L(k) = \dot{R}_L(k) + v(k) \quad (3)$$

The lead HV acceleration $a_L$ is estimated by taking the derivative of $v_L$ using Forward Euler Approximation [41]. A moving average filter with a window size 16 is used to smooth the estimated $a_L$. The velocity and acceleration in an example car-following event are shown in Fig. 4.

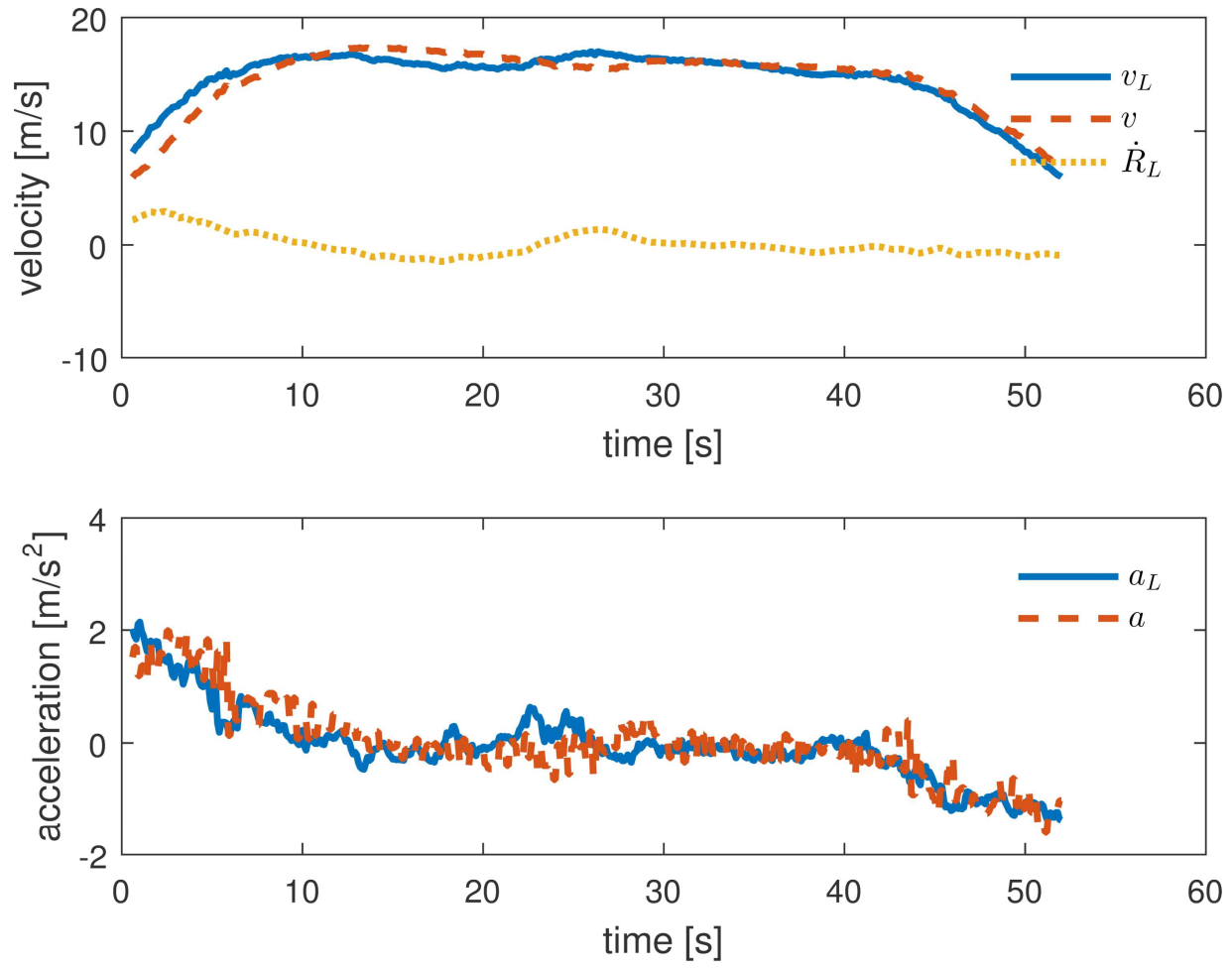

Fig. 4. Estimation of the lead vehicle acceleration

The driver model parameter vector $\boldsymbol{h}$ is estimated based on the Weighted Least Square Method [42], which is less influenced by the outliers than the standard least-squares fit [43], [44]. Define the acceleration vector of the lead vehicle

$$a_L^{(n_s)}(i{:}j) = \left[a_L^{(n_s)}(i), a_L^{(n_s)}(i+1), \dots, a_L^{(n_s)}(j)\right] \quad (4)$$

with $n_s = 1,2, \dots, N_s$ and $j \geq i$, where $a_L^{(n_s)}(i)$ prepresents the $i^{\text{th}}$ step acceleration of the HV in the $n_s{}^{\text{th}}$ car-following sequence. $N_s$ is the total number of car-following sequences in the SPMD database. ":" denotes continuous integer indexes, i.e., $x(i{:}j) = [x(i), x(i+1), \dots, x(j)]^T$ with $j \geq i$. Let

$$\boldsymbol{y_h} = \left[a_L^{(1)}(2{:}N_h^{(1)}), a_L^{(2)}(2{:}N_h^{(2)}), \dots, a_L^{(N_s)}\left(2{:}N_h^{(N_s)}\right)\right]^T \quad (5)$$

where $N_h^{(n_s)}$ represents the size of the acceleration vector. Define the input vector of each car-following scenario as

$$x_p^{(n_s)} = \begin{bmatrix} 1 & a_L^{(n_s)}(1) & v_L^{(n_s)}(1) \\ 1 & a_L^{(n_s)}(2) & v_L^{(n_s)}(2) \\ \vdots & \vdots & \vdots \\ 1 & a_L^{(n_s)}\left(N_h^{(n_s)}-1\right) & v_L^{(n_s)}\left(N_h^{(n_s)}-1\right) \end{bmatrix}^T \quad (6)$$

The input vector of $N_s$ car-following events is defined as

$$\boldsymbol{\chi_h} = \left[x_h^{(1)}, x_h^{(2)}, \dots, x_h^{(n_s)}\right]^T \quad (7)$$

The Matlab® function *robustfit* [43] is used to fit $\boldsymbol{h}$ with input vector $\boldsymbol{\chi_h}$ and observer vector $\boldsymbol{y_h}$ based on the "bisquare" approach [42]. The standard deviation of $u_h$ is calculated from the estimation error $(a_L(k+1) - [1, a_L(k), v_L(k)]\,\boldsymbol{h})$.

### C. Automated Vehicle Model

The AV model consists of two parts: the longitudinal vehicle dynamic and the control system.

#### 1) Vehicle Dynamics

We use a longitudinal vehicle dynamic model from [45]

$$M\frac{dv(t)}{dt} = F_x(t) - Mg\,sin\theta_{rg}(t) - f_{rr}Mg\,cos\theta_{rg}(t) - 0.5\rho_{air}A_vC_d\big(v(t)+v_w(t)\big)^2 \quad (8)$$

where $M$ is the vehicle mass, $F_x$ is the longitudinal force, $\theta_{rg}$is the road grade angle, g is the gravitational constant, $f_{rr}$ is the rolling resistance coefficient, $\rho_{air}$ is the air density, $A_v$ is the frontal area of AV, $C_d$ is the aerodynamic coefficient and $v_w$ is the wind speed.

At equilibrium (i.e. when $dv/dt = 0$),

$$F_{x0}(t) = Mg\,sin\theta_{rg}(t) + f_{rr}Mg\,cos\theta_{rg}(t) + 0.5\rho_{air}A_vC_d\big(v_0(t)+v_w(t)\big)^2 \quad (9)$$

It can be linearized around the equilibrium point by using the Taylor series expansion

$$\tau\frac{d\tilde{v}}{dt} + \tilde{v} = K_{AV}(\tilde{F}_x + d_{AV}) \quad (10)$$

where $\tilde{v}$ is the velocity deviation, defined as

$$\tilde{v} = v - v_0 \quad (11)$$

and $\tilde{F}_x$ is the longitudinal force deviation, defined as

$$\tilde{F}_x = F_x - F_{x0} \quad (12)$$

Parameters $\tau$ , $K_{AV}$ , and $d_{AV}$ were derived as $\tau = M/(\rho_{air}C_dA_v(v_0+v_w))$ , $K_{AV} = 1/\rho_{air}C_dA_v(v_0+v_w)$ , and $d_{AV} = Mg\big(f_{rr}sin\theta_{rg} - cos\theta_{rg}\big)\mathrm{d}\theta_{rg}/\mathrm{d}t$. Assume $v_w = 0$ and $\theta_{rg} = 0$. Using Laplace transformation [46] on (10), we obtain a first order lag system representing the vehicle longitudinal dynamics.

$$\frac{\tilde{v}(s)}{\tilde{F}_x(s)} = \frac{K_{AV}}{\tau s + 1} \quad (13)$$

#### 2) Longitudinal Control

The longitudinal control is designed to follow the lead HV velocity and to maintain a proper distance [47]. As shown in Fig. 5, the control algorithm can be expressed as

$$\tilde{F}_x(z) = \left(K_p + K_i\frac{T_s}{z-1}\right)\tilde{R}_L(z) + K_d\dot{R}_L(z) \quad (14)$$

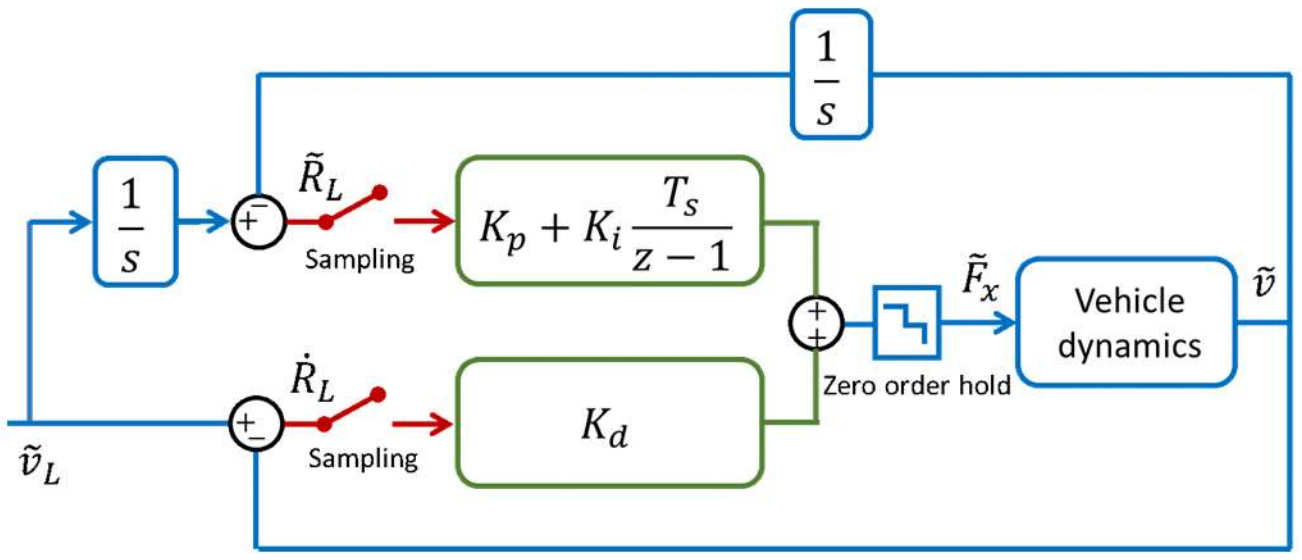

Fig. 5. Automated vehicle control model

where $F_x(z)$, $\tilde{R}_L(z)$, and $\dot{R}_L(z)$ are $Z$ transformation [48] of $F_x(t)$, $\tilde{R}_L(t)$, and $\dot{R}_L(t)$. $T_s$ is the time step. $\tilde{R}_L(t)$ is defined as

$$\tilde{R}_L(t) = R_L(t) - R_L^{desire} \quad (15)$$

where $R_L^{desire} = v_0 * t_{HW}^{desire}$ and $t_{HW}^{desire}$ is the desired time headway. $\tilde{R}_L$ is regulated by a discretized Proportional-Integral (PI) controller and the range rate is regulated by a Proportional (P) controller. $\tilde{R}_L$ is calculated from

$$\tilde{R}_L(t) = \int_0^t \dot{R}_L(t)dt + R_{L0} - R_L^{desire} \quad (16)$$

where $R_{L0}$ is the initial range. $R_{L0}$ is equal to $R_L^{desire}$ to make the test starts from an equilibrium.

## III. Accelerated Evaluation

The statistics of the lead HV motion are modified to generate more intense maneuvers. From (1), the acceleration of the lead HV follows the probability density function

$$a_L(k+1) \sim \mathcal{N}(h_0 + h_1a_L(k) + h_2v_L(k), \sigma_u^2) \quad (17)$$

The general idea behind accelerating the evaluation is to add a series of biases $[\mathscr{b}(1), \mathscr{b}(2), \dots]$ to the mean of the acceleration distribution at each step. The modified acceleration distribution then becomes

$$a_L(k+1) \sim \mathcal{N}(h_0 + h_1a_L(k) + h_2v_L(k) + \mathscr{b}(k), \sigma_u^2) \quad (18)$$

In this section, we first discretize the car-following model and rewrite it in the state space. We then calculate the optimal $\mathscr{b}(k)$ using a stochastic optimization approach to find the key test scenarios. Importance Sampling theory is used to interpret the results in the accelerated tests and estimate the real world performance.

### A. State Space Form of the Car-Following Model

In discrete time, the lead vehicle velocity is calculated from

$$v_L(k+1) = v_L(k) + T_sa_L(k) \quad (19)$$

Define velocity deviation of the lead HV as

$$\tilde{v}_L(k) = v_L(k) - v_0 \quad (20)$$

We have

$$\tilde{v}_L(k+1) = \tilde{v}_L(k) + T_sa_L(k) \quad (21)$$

Define

$$u = u_h + h_0 + h_2v_0 \quad (22)$$

From (1) and (20), we have

$$a_L(k+1) = h_1a_L(k) + h_2\tilde{v}_L(k) + u \quad (23)$$

Discretize (13) using the Zero Order Hold method [48],

$$\frac{\tilde{v}(z)}{\tilde{F}_x(z)} = \frac{n_v z^{-1}}{1 + d_v z^{-1}} \quad (24)$$

where $\tilde{v}(z)$ and $\tilde{F}_x(z)$ are the $Z$ transformation of $\tilde{v}(t)$ and $\tilde{F}_x(t)$. $n_v$ and $d_v$ are coefficients of the discrete system, which can be directly calculated using Matlab® function *c2d*. Taking the inverse $Z$ transformation of (24), we get

$$\tilde{v}(k+1) = -d_v \tilde{v}(k) + n_v \tilde{F}_x(k) \quad (25)$$

Substitute (11) and (20) into (3).

$$\dot{R}_L(k) = \tilde{v}_L(k) - \tilde{v}(k) \quad (26)$$

Substitute (15) and (19) into (16).

$$\begin{aligned} \tilde{R}_L(k+1) &= \tilde{R}_L(k) + T_s \dot{R}_L(k) \\ &= \tilde{R}_L(k) + T_s \tilde{v}_L(k) - T_s \tilde{v}\,(k) \end{aligned} \quad (27)$$

Take the inverse $Z$ transformation of (14).

$$\begin{aligned} \tilde{F}_x(k+1) &= \tilde{F}_x(k) + K_p \tilde{R}_L(k+1) \\ &+ (-K_p + K_i T_s)\tilde{R}_L(k) + K_d \dot{R}_L(k+1) - K_d \dot{R}_L(k) \end{aligned} \quad (28)$$

Substitute (21), (25), (26) and (27) into (28).

$$\begin{aligned} \tilde{F}_x(k+1) = q_1 a_L(k) + q_2 \tilde{v}_L(k) + q_3 \tilde{v}(k) + q_4 \tilde{F}_x(k) \\ + q_5 \tilde{R}_L(k) \end{aligned} \quad (29)$$

where $q_1 = K_d T_s$, $q_2 = K_p T_s$, $q_3 = K_d + K_d d_v - K_p T_s$, $q_4 = 1 - K_d n_v$, $q_5 = K_i T_s$.

Rewriting (21), (23), (25), (27), and (29) into the state space form, we have

$$\begin{aligned} X(k+1) &= AX(k) + Bu(k) \\ R_L(k) &= CX(k) \end{aligned} \quad (30)$$

where $X(k) = \begin{bmatrix} a_L(k) & \tilde{v}_L(k) & \tilde{v}(k) & \tilde{F}_x(k) & \tilde{R}_L(k) \end{bmatrix}^T$

$$A = \begin{bmatrix} h_1 & h_2 & 0 & 0 & 0 \\ T_s & 1 & 0 & 0 & 0 \\ 0 & 0 & -d_v & n_v & 0 \\ q_1 & q_2 & q_3 & q_4 & q_5 \\ 0 & T_s & -T_s & 0 & 1 \end{bmatrix}$$

$$B = [1 \quad 0 \quad 0 \quad 0 \quad 0]^T, \quad C = [0 \quad 0 \quad 0 \quad 0 \quad 1]$$

with initial condition $X(1) = [0 \quad 0 \quad 0 \quad 0 \quad 0\,]^T$ and physical constraints

$$X_{min} \le X(k) \le X_{max}$$

where $X_{min} = [a_L^{Min}, v_L^{Min} - v_0, v^{Min} - v_0, F_x^{Min} - F_{x0}, R_L^{Min} - R_L^{desire}]^T$ and $X_{max} = [a_L^{Max}, v_L^{Max} - v_0, v^{Max} - v_0, F_x^{Max} - F_{x0}, R_L^{Max} - R_L^{desire}]^T$.

## B. Accelerated Evaluation

The Accelerated Evaluation procedure follows four steps: 1) calculate the optimal $\mathscr{b}(k)$, 2) randomize the termination time, 3) simulate using the new statistical model, and 4) estimate the real world benefits. The same procedure was applied to estimate three safety metrics: crash, injury, and conflict.

Define the rare event of interest $\mathcal{E}$ as

$$\mathcal{E} = \{\min(R_L(k)) < R_{\mathcal{E}} | 1 \le k \le K\} \quad (31)$$

where $R_{\mathcal{E}}$ is the critical distance. For crash and injury, $R_{\mathcal{E}} = 0$. For conflict, $R_{\mathcal{E}} = 30$ feet representing the approximating zone. $K$ is the maximum number of time steps. The indicator function of $\mathcal{E}$ is defined as

$$I_{\mathcal{E}}(n) = \begin{cases} 1, & \mathcal{E}\ happens \\ 0, & otherwise \end{cases} \quad (32)$$

Our goal is to estimate the probability of $\mathcal{E}$ occurring, i.e.

$$\gamma = \mathbb{P}(\mathcal{E}) = \mathbb{E}(I_{\mathcal{E}}) \quad (33)$$

### 1) Calculation of the Optimal Mean Shift

The optimal value of $\mathscr{b}(k)$, denoted as $\mathscr{b}^*(k)$, is calculated in two steps. First we calculate $u^*(k)$ - the optimal realization of $u(k)$. Second, $\mathscr{b}^*(k)$ is calculated to maximize the likelihood for $u(k)$ to be $u^*(k)$ in the accelerated tests.

From (1) and (22), under the naturalistic driving conditions, we have

$$u(k) \sim f_{u_k}(u(k)) = \mathcal{N}(\mu_u, \sigma_u^2) \quad (34)$$

where $\mu_u = h_0 + h_2 v_0$. The simulation ends when either $\mathcal{E}$ occurs or the maximum time step is reached. Define the termination time step

$$k_{\mathcal{T}} = \min\{min(k | R_L(k) < R_{\mathcal{E}}), K\} \quad (35)$$

$k_{\mathcal{T}}$ is an integer $\in \{1, \ldots, K\}$. Because the lead HV is modeled as a Markov Chain, the probabilistic density distribution for a car-following event (from 1 to $k_{\mathcal{T}}$) can be calculated from

$$f_u(u(1{:}\,k_{\mathcal{T}} - 1)) = \prod_{k=1}^{k_{\mathcal{T}}-1} f_{u_k}(u(k)) \quad (36)$$

Substituting (34) into (36), we have

$$\begin{aligned} & f_u(u(1{:}\,k_{\mathcal{T}} - 1)) \\ &= \left(\frac{1}{\sqrt{2\pi}\sigma_u}\right)^{k_{\mathcal{T}}-1} exp\left(-\frac{\|u(1{:}\,k_{\mathcal{T}} - 1) - \mu_u \mathbf{1}\|_2^2}{2\sigma_u^2}\right) \end{aligned} \quad (37)$$

where $u(1{:}\,k_{\mathcal{T}} - 1) = [u(1), u(2), \ldots, u(k_{\mathcal{T}} - 1)]^T$, $\|\cdot\|_2^2$ represents the Euclidean norm and $\mathbf{1} = [1,1,\ldots,1]^T \in \mathbb{R}^{(k_{\mathcal{T}}-1)\times 1}$.

In the accelerated tests, from (18) and (22), we have

$$u(k) \sim \tilde{f}_{u_k}(u(k)) = \mathcal{N}(\mu_u + \mathscr{b}(k), \sigma_u^2) \quad (38)$$

The modified probabilistic density distribution is calculated from

$$\tilde{f}_u(u(1{:}\,k_{\mathcal{T}} - 1)) = \prod_{k=1}^{k_{\mathcal{T}}-1} \tilde{f}_{u_k}(u(k)) \quad (39)$$

Substituting (38) into (39), we have

$$\begin{aligned} \tilde{f}_u(u(1{:}\,k_{\mathcal{T}} - 1)) &= \left(\frac{1}{\sqrt{2\pi}\sigma_u}\right)^{k_{\mathcal{T}}-1} \\ \exp&\left(-\frac{\|u(1{:}\,k_{\mathcal{T}} - 1) - \mathscr{b}(1{:}\,k_{\mathcal{T}} - 1) - \mu_u \mathbf{1}\|_2^2}{2\sigma_u^2}\right) \end{aligned} \quad (40)$$

The optimal realization of $u(k)$ is calculated to achieve the following two goals:

i) To make $\mathcal{E}$ occur at a time $k_{\mathcal{T}}^*$
ii) To maximize the likelihood the occurrence of $\mathcal{E}$

We need to i) calculate the probability of $\mathcal{E}$, and ii) focus on the events with a high probability of occurrence, since $\mathcal{E}$ occurring at an extremely low probability will have negligible effects in calculating $\mathbb{P}(\mathcal{E})$. The two goals can be expressed as a stochastic optimization problem shown as follows.

$$u_{k_T^*}(1:k_T^*-1)=\arg\max_{u(1:k_T^*-1)} f_u\big(u(1:k_T^*-1)\big)$$

subject to

$$\begin{gathered} R_L(k_T^*)\le R_{\mathcal{E}} \\ \boldsymbol{X_{min}}\le \boldsymbol{X}(k)\le \boldsymbol{X_{max}} \\ u_{min}\le u(k)\le u_{max} \end{gathered} \tag{41}$$

for $k=1,2,\dots,k_T^*-1$

Substitute (36) and (30) into (41). The optimization problem can be rewritten in a quadratic programming form.

$$u_{k_T^*}(1:k_T^*-1)=\arg\min_{u(1:k_T^*-1)}\left[\frac{1}{2}u(1:k_T^*-1)^T u(1:k_T^*-1)-\mu_u\mathbf{1}^T u(1:k_T^*-1)\right] \tag{42}$$

subject to

$$\boldsymbol{A_{k_T^*}}\,u_{k_T^*}(1:k_T^*-1)\le \boldsymbol{b_{k_T^*}}$$

where

$$\boldsymbol{A_{k_T^*}}=\begin{bmatrix} \boldsymbol{CA}^{(k_T^*-2)}\boldsymbol{B} & \boldsymbol{CA}^{(k_T^*-3)}\boldsymbol{B} & \cdots & \boldsymbol{CB} \\ \boldsymbol{B} & 0 & \cdots & 0 \\ \boldsymbol{AB} & \boldsymbol{B} & \cdots & 0 \\ \vdots & \vdots & \ddots & \vdots \\ \boldsymbol{A}^{(k_T^*-3)}\boldsymbol{B} & \boldsymbol{A}^{(k_T^*-4)}\boldsymbol{B} & \cdots & 0 \\ -\boldsymbol{B} & 0 & \cdots & 0 \\ -\boldsymbol{AB} & -\boldsymbol{B} & \cdots & 0 \\ \vdots & \vdots & \ddots & \vdots \\ -\boldsymbol{A}^{(k_T^*-3)}\boldsymbol{B} & -\boldsymbol{A}^{(k_T^*-4)}\boldsymbol{B} & \cdots & 0 \\ 1 & 0 & \cdots & 0 \\ 0 & 1 & \cdots & 0 \\ \vdots & \vdots & \ddots & \vdots \\ 0 & 0 & \cdots & 1 \\ -1 & 0 & \cdots & 0 \\ 0 & -1 & \cdots & 0 \\ \vdots & \vdots & \ddots & \vdots \\ 0 & 0 & \cdots & -1 \end{bmatrix},\ \boldsymbol{b_{k_T^*}}=\begin{bmatrix} R_{\mathcal{E}}-\boldsymbol{CA}^{(k_T^*-1)}\boldsymbol{X}(1) \\ \boldsymbol{X_{max}}-\boldsymbol{AX}(1) \\ \boldsymbol{X_{max}}-A^2X(1) \\ \vdots \\ \boldsymbol{X_{max}}-\mathbf{A}^{(k_T^*-2)}\boldsymbol{X}(1) \\ -\boldsymbol{X_{min}}+\boldsymbol{AX}(1) \\ -\boldsymbol{X_{min}}+\boldsymbol{A}^2\boldsymbol{X}(1) \\ \vdots \\ -\boldsymbol{X}_{\mathbf{min}}+-\mathbf{A}^{(k_T^*-2)}\boldsymbol{X}(1) \\ u_{max} \\ u_{max} \\ \vdots \\ u_{max} \\ -u_{min} \\ -u_{min} \\ \vdots \\ -u_{\min} \end{bmatrix}$$

The optimal shift parameters $\mathscr{b}_{k_T^*}(1:K)$ are calculated to maximize the likelihood of $u_{k_T^*}(1:K)$. For a specific $k_T^*$, $\mathscr{b}_{k_T^*}(1:K)$ can be calculated from

$$\mathscr{b}_{k_T^*}(1:k_T^*)=\arg\max_{\mathscr{b}(1:\mathrm{K})}\tilde{f}_u\left(u_{k_T^*}(1:k_T^*),\mathscr{b}(1:k_T^*)\right) \tag{43}$$

Substituting (40) into (43), we have

$$\mathscr{b}_{k_T^*}(1:k_T^*)=\arg\max_{\mathscr{b}(1:k_T^*)}\left(\frac{1}{\sqrt{2\pi}\sigma_u}\right)^{k_T^*}\exp\left(-\frac{\|u(1:k_T^*)-\mathscr{b}(1:k_T^*)-\mu_u\mathbf{1}\|_2^2}{2\sigma_u^2}\right) \tag{44}$$

Take out the constant terms from the $\arg max$ function.

$$\mathscr{b}_{k_T^*}(1:k_T^*)=\arg\max_{\mathscr{b}(1:k_T^*)}\exp(-\|u(1:k_T^*)-\mathscr{b}(1:k_T^*)-\mu_u\mathbf{1}\|_2^2) \tag{45}$$

Because the exponential function is monotonically increasing, it will not affect the solution of the $\arg max$ function. $\mathscr{b}_{k_T^*}(1:k_T^*)$ can be calculated from

$$\mathscr{b}_{k_T^*}(1:k_T^*)=\arg\min_{\mathscr{b}(1:k_T^*)}\|u(1:k_T^*)-\mathscr{b}(1:k_T^*)-\mu_u\mathbf{1}\|_2^2 \tag{46}$$

Since $\|u(1:k_T^*)-\mathscr{b}(1:k_T^*)-\mu_u\mathbf{1}\|_2^2\ge 0$, to achieve the minimum value, we have

$$\|u(1:k_T^*)-\mathscr{b}(1:k_T^*)-\mu_u\mathbf{1}\|_2^2=0 \tag{47}$$

Therefore, the optimal mean shift is calculated from

$$\mathscr{b}_{k_T^*}(1:k_T^*)=u_{k_T^*}(1:k_T^*)-\mu_u\mathbf{1} \tag{48}$$

$\mathscr{b}_{k_T^*}(1:K)$ for each $k_T^*=1,\dots,K$ will be calculated offline. We will use these $\mathscr{b}_{k_T^*}$ in the stochastic process during the evaluation.

*2) Randomization of Termination Time*

The termination time $k_T^*$ is fixed in the previous section. However, $\mathcal{E}$ (e.g., a crash) may occur at any moment before time step $K$. $k_T^*$ needs to be randomized to reflect all possible scenarios.

We first set the boundary of $k_T^*$. When $k_T^*$ is small, it may fail to find a feasible solution to the optimization problem in (42) because no physically possible HV motion will lead to a crash when the test duration is very short. Let $k_{Tmin}^*$ be the minimum value of $k_T^*$ that has a feasible solution in (42) under physical constraints.

$$\begin{gathered} k_{Tmin}^*=\min k_T^* \\ \text{such that } \exists\ u_{k_T^*}(1:k_T^*-1) \text{ that makes} \\ \boldsymbol{A_{k_T^*}}\,u_{k_T^*}(1:k_T^*-1)\le \boldsymbol{b_{k_T^*}} \end{gathered} \tag{49}$$

$k_T^*$ is then randomized using a uniform distribution

$$k_T^*\sim f_{k_T^*}(k_T^*)=\frac{1}{K-k_{Tmin}^*+1} \tag{50}$$

where $k_T^*=k_{Tmin}^*,k_{Tmin}^*+1,\dots,K-1,K$.

It should be noted that $k_T^*$ is different from $k_T$. $k_T$ is the real termination time in the simulation, while $k_T^*$ is used to randomize termination time in calculating the optimal shift $\mathscr{b}_{k_T^*}(1:k_T^*-1)$.

*3) Conducting the Accelerated Tests*

The car-following simulation is executed with the lead HV following the modified distribution.

$$a_L(k+1)\sim\mathcal{N}(h_0+h_1a_L(k)+h_2v_L(k)+\mathscr{b}_{k_T^*}(k),\sigma_u^2) \tag{51}$$

Record whether $\mathcal{E}$ occurs, and if it does, the termination time $k_T$ defined in (35).

*4) Calculation of Safety Benefit in the Real World*

The real-world safety benefit is calculated based on the Importance Sampling (IS) theory. By modifying the HV statistics from (17) to (18), more intense HV maneuvers are generated. Using a different distribution, however, leads to biased samples, and the key to IS is to compensate for this bias and to compute the safety performance of AVs in the real world. IS has been successfully applied to analyze critical events in reliability [49], finance [50], insurance [51], and telecommunication networks [52]. General reviews of IS can be found in [53]–[55]. To the best of our knowledge, no prior research has applied IS to AV evaluation in the car-following scenario.

We describe the core mechanism of IS as follows. First, the likelihood ratio L in the $n^{th}$ car following event is defined as

$$L(n)=\frac{f\left(a_L(1:k_{\mathcal{T}}-1)\right)}{f^*\left(a_L(1:k_{\mathcal{T}}-1)\right)} \tag{52}$$

which is the ratio between $f(\cdot)$ the likelihood in the naturalistic driving condition and $f^*(\cdot)$ the likelihood in the accelerated tests.

Based on (17), $f\left(a_L(1:k_{\mathcal{T}}-1)\right)$ can be calculated from

$$f\left(a_L(1:k_{\mathcal{T}}-1)\right)= \prod_{k=1}^{k_{\mathcal{T}}-1}\frac{1}{\sigma_u\sqrt{2\pi}}exp\left\{-\frac{\left(a_L(k+1)-h_1a_L(k)-h_2v_L(k)-h_0\right)^2}{2\sigma_u^2}\right\} \tag{53}$$

$f^*\left(a_L(1:k_{\mathcal{T}}-1)\right)$ is calculated from

$$f^*\left(a_L(1:k_{\mathcal{T}}-1)\right)= \sum_{k_{\mathcal{T}}^*=k_{\mathcal{T}min}^*}^{K}\tilde{f}\left(a_L(1:k_{\mathcal{T}}-1)|k_{\mathcal{T}}^*\right)f_{k_{\mathcal{T}}^*}(k_{\mathcal{T}}^*) \tag{54}$$

Based on (18), $\tilde{f}\left(a_L(1:k_{\mathcal{T}}-1)|k_{\mathcal{T}}^*\right)$ is calculated from

$$\tilde{f}\left(a_L(1:k_{\mathcal{T}}-1)|k_{\mathcal{T}}^*\right)=\prod_{k=1}^{k_{\mathcal{T}}-1}\frac{1}{\sigma_u\sqrt{2\pi}} exp\left\{-\frac{\left(a_L(k+1)-h_1a_L(k)-h_2v_L(k)-h_0-\mathscr{b}_{k_{\mathcal{T}}^*}(k)\right)^2}{2\sigma_u^2}\right\} \tag{55}$$

For notational simplicity, denote $\boldsymbol{x}$ as the random vector describing the motions of the lead HV i.e. $\boldsymbol{x}=a_L(1:k_{\mathcal{T}}-1)$. The probability of $\mathcal{E}$ satisfies

$$\begin{aligned}\mathbb{P}(\mathcal{E})&=\mathbb{E}_f\left(I_{\mathcal{E}}(\boldsymbol{x})\right)=\int I_{\mathcal{E}}(\boldsymbol{x})f(\boldsymbol{x})d\boldsymbol{x}\\&=\int[I_{\mathcal{E}}(\boldsymbol{x})L(\boldsymbol{x})]f^*(\boldsymbol{x})\mathrm{d}\boldsymbol{x}=\mathbb{E}_{f^*}\left(I_{\mathcal{E}}(\boldsymbol{x})L(\boldsymbol{x})\right)\end{aligned} \tag{56}$$

The overall IS estimator of $\gamma$ in $n$ tests is then calculated from

$$\hat{\gamma}_n=\frac{1}{n}\sum_{i=0}^{n}I_{\mathcal{E}}(\boldsymbol{x}_i)L(\boldsymbol{x}_i) \tag{57}$$

The accuracy of the estimation is represented by the relative half-width, which is the half-width of the confidence interval relative to the probability to be estimated. With the Confidence Level at $100(1-\alpha)\%$, the relative half-width of $\hat{\gamma}_n$ is defined as

$$l_r(n)=\frac{l_\alpha(n)}{\hat{\gamma}_n} \tag{58}$$

$l_\alpha$, the half-width, is defined as

$$l_\alpha=z_\alpha\sigma(\hat{\gamma}_{1:n}) \tag{59}$$

where $\sigma(\hat{\gamma}_{1:n})$ represents the standard deviation of $[\hat{\gamma}_1,\hat{\gamma}_2,\dots,\hat{\gamma}_n]^T$. $z_\alpha$ is defined by

$$z_\alpha=\Phi^{-1}(1-\alpha/2) \tag{60}$$

where $\Phi^{-1}$ is the inverse cumulative distribution function of $\mathcal{N}(0,1)$. The Central Limit Theorem [56] implies that, as more tests are conducted, $\hat{\gamma}_n$ will converge to $\gamma$ and $l_r$ will reduce to zero. The convergence is reached as $l_r$ is smaller than constant threshold $\beta$.

The procedure of the accelerated test is summarized in Fig. 6.

*Before simulations*:

i) Calculate the optimal mean shift vector $\mathscr{b}_{k_{\mathcal{T}}^*}(1:K)$ for $k_{\mathcal{T}}^*=1,2,\dots,K$ from (48).
ii) $k_{\mathcal{T}min}^*$ is calculated from (49). Generate the distribution $f_{k_{\mathcal{T}}^*}(\cdot)$ from (50) using $k_{\mathcal{T}min}^*$.

*During simulations*:

iii) Sample a $k_{\mathcal{T}}^*$ from (50).
iv) Run the accelerated tests with AE distribution in (51). After each test record indicator function $I_{\mathcal{E}}$ and the termination time $k_{\mathcal{T}}$ from (32) and (35).
v) Calculate the likelihood of each test from (52).
vi) Calculate the $\hat{\gamma}_n$ from (57).
vii) Calculate the relative half-width $l_r$ in (58).
viii) If $l_r<\beta$, output $\hat{\gamma}_n$. Otherwise, go back to step iii)

Note that we have chosen to modify the mean in our IS algorithm rather than other parameters such as the standard deviation of the Gaussian distributions. This choice is motivated by the fact that, in this case, the likelihood ratio evaluated under the rare event, which is the output of the simulation, turns out to have a half-width that grows at most polynomially as the rarity of the event increases. This phenomenon, in turn, keeps the required sample size under control to achieve a reasonable estimation accuracy. Research in [57], [58] provide general studies on such efficiency considerations. Moreover, a good choice of the modified mean can be found via a convex optimization, which can be readily solved with available numerical routines. We also note that the idea of randomizing the termination time has appeared in the rare-event simulation literature [59], [60].

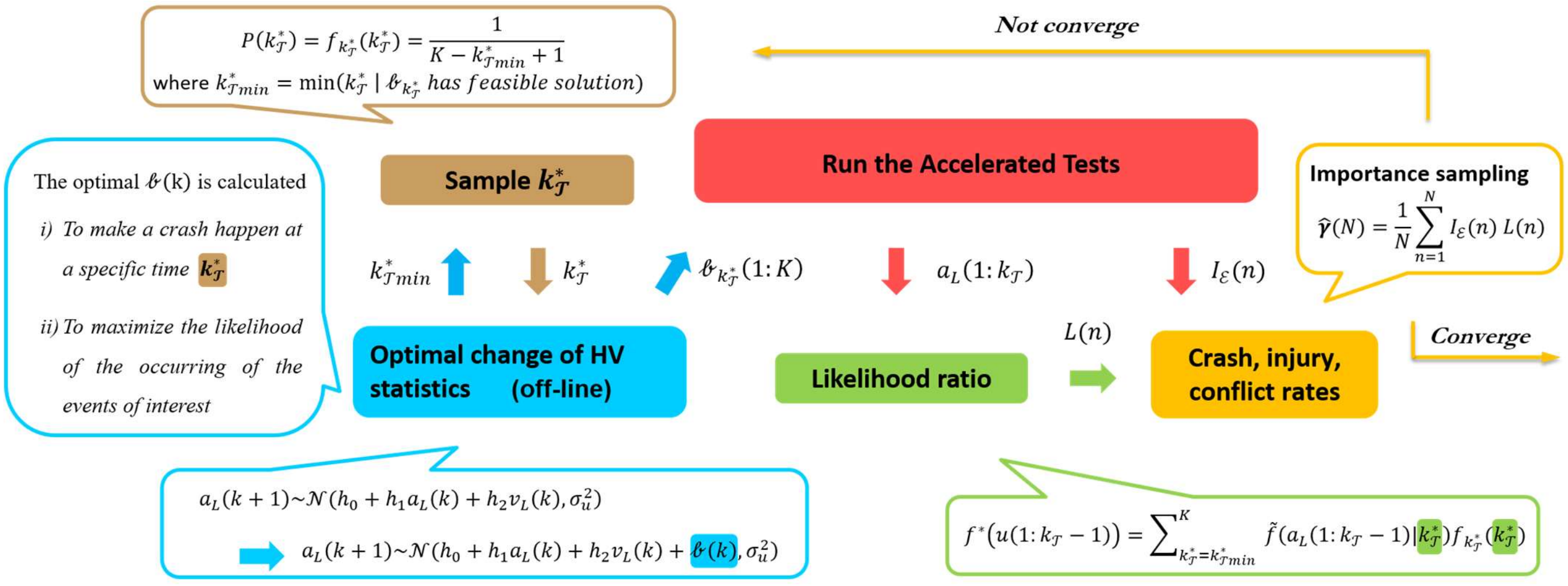

Fig. 6. Procedure to calculate crash rate in the car-following scenario

## IV. Results and Analysis

The uniform distribution was used as the benchmark for modified distribution in the accelerated tests. The behavior of the HV was modified to be more aggressive, but the evaluation process may or may not accelerate. Three metrics—crash, injury, and conflict rates—were calculated. In each case, both accelerated and naturalistic driving simulations (non-accelerated, based on Monte Carlo) were conducted to examine the accuracy and the accelerated rates of the proposed method.

### A. Simulation Results with uniform distribution

The uniform AE distribution is used as the baseline for comparison. The HV acceleration is generated from

$$a_L(k+1) \sim \tilde{f}_{ud}\big(a_L(k), v_L(k)\big)$$
$$= \mathcal{U}\left([1, a_L(k), v_L(k)]\boldsymbol{h} - \frac{\vartheta_{ud}}{2}, [1, a_L(k), v_L(k)]\boldsymbol{h} + \frac{\vartheta_{ud}}{2}\right) \quad (61)$$

where $\mathcal{U}$ represents the uniform distribution and parameter $\vartheta_{ud}$ is chosen to be $6\sigma_u$.

The likelihood ratio of the uniform distribution method can be calculated from

$$L(n) = \frac{f\big(a_L(1{:}k_\mathcal{T}-1)\big)}{f_{ud}^*\big(a_L(1{:}k_\mathcal{T}-1)\big)} \quad (62)$$

where $k_\mathcal{T}$ is the termination time defined in (35). $f\big(a_L(1{:}k_\mathcal{T}-1)\big)$ is calculated from (53). $f_{ud}^*(\cdot)$ is calculated from

$$f_{ud}^*\big(a_L(1{:}k_\mathcal{T}-1)\big) = \prod_{k=1}^{k_\mathcal{T}-1} \tilde{f}_{ud}\big(a_L(k)\big) = \vartheta_{ud}^{-k_\mathcal{T}} \quad (63)$$

The model parameters used in the simulations are listed in TABLE II. A sample simulation run is shown in Fig. 7. Although the lead HV acceleration oscillates up and down,

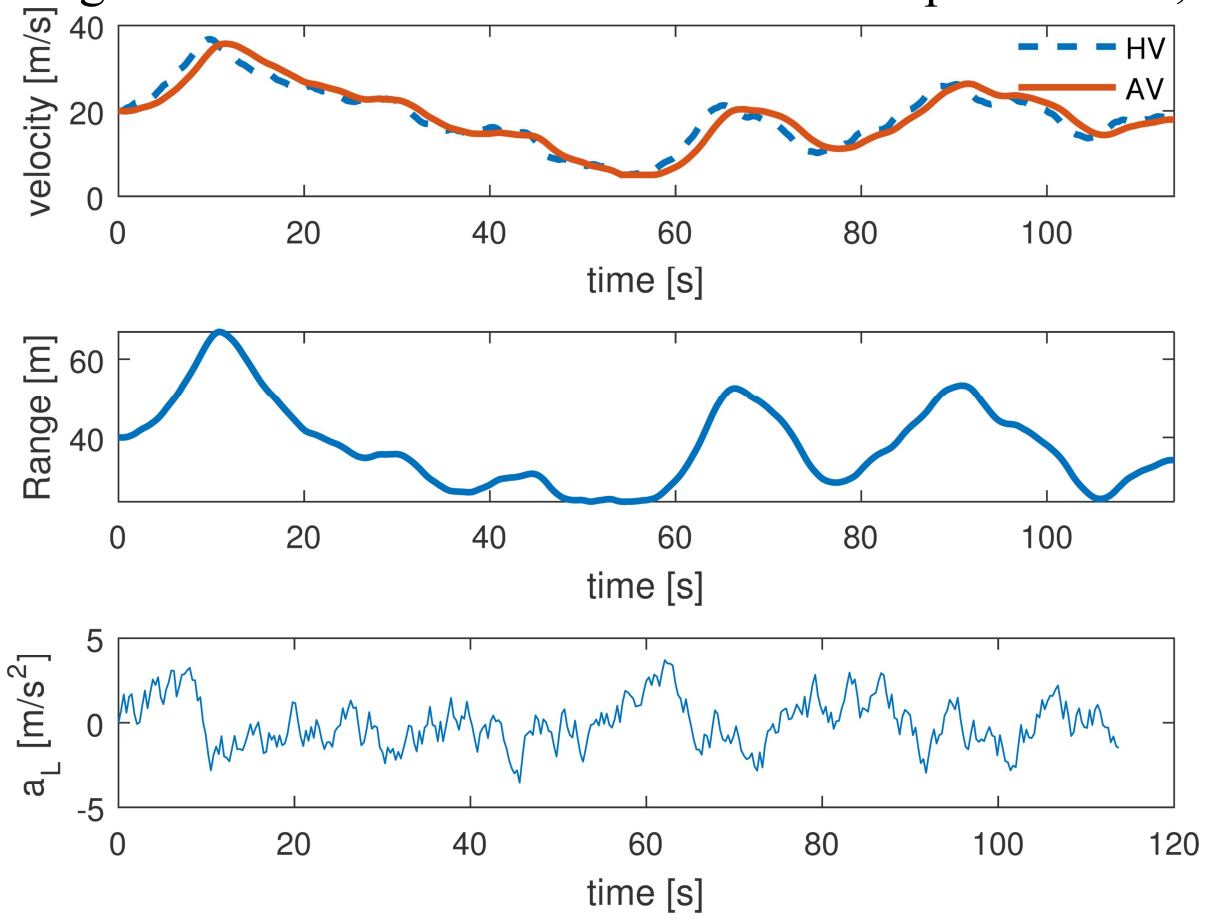

Fig. 7. An example maneuver generated by the baseline accelerated evaluation approach using uniform distribution

TABLE II
Parameters Used in the Car-Following Simulation

| Var. | Unit | Value | Var. | Unit | Value |
|---|---|---|---|---|---|
| $a_{L_0}$ | m/s$^2$ | 0 | $K_p$ | - | 62.63 |
| $a_0$ | m/s$^2$ | 0 | $M$ | kg | 1757 |
| $a_L^{Max}$ | m/s$^2$ | 9.81 | $R_L^{Max}$ | m | 1e3 |
| $a_L^{Min}$ | m/s$^2$ | -9.81 | $R_L^{Min}$ | m | 0 |
| $a^{Max}$ | m/s$^2$ | 9.81 | $T_{CF}$ | s | 114 |
| $a^{Min}$ | m/s$^2$ | -9.81 | $t_{HW}^{desire}$ | s | 2 |
| $A_v$ | m$^2$ | 2.2 | $T_s$ | s | 0.3 |
| $C_d$ | - | 0.32 | $u_{max}$ | m/s$^2$ | 1.2 |
| $F_x^{Max}$ | N | 17236 | $u_{min}$ | m/s$^2$ | -1.2 |
| $F_x^{Min}$ | N | -17236 | $v^{Max}$ | m/s | 50 |
| $g$ | m/s$^2$ | 9.81 | $v^{Min}$ | m/s | 1 |
| $h_0$ | - | 3.395e-2 | $v_L^{Max}$ | m/s | 50 |
| $h_1$ | - | 0.8516 | $v_L^{Min}$ | m/s | 1 |
| $h_2$ | - | -1.406e-3 | $v_0$ | m/s | 20 |
| $K$ | - | 119 | $v_{L_0}$ | m/s | 20 |
| $K_d$ | - | 882.7 | $\rho_{air}$ | kg/m$^3$ | 1.202 |
| $K_i$ | - | 1.111 | $\sigma_u$ | - | 0.3949 |

overall the lead vehicle motion is not hard to follow. Therefore, a crash rarely occurs.

The estimated crash rate is plotted in Fig. 8. A million simulations were conducted, but because the convergence is very weak, not much can be said about the crash rate of the AV. The uniform distribution did not effectively accelerate crash evaluation because it did not consider the correlations between samples in a dynamic system.

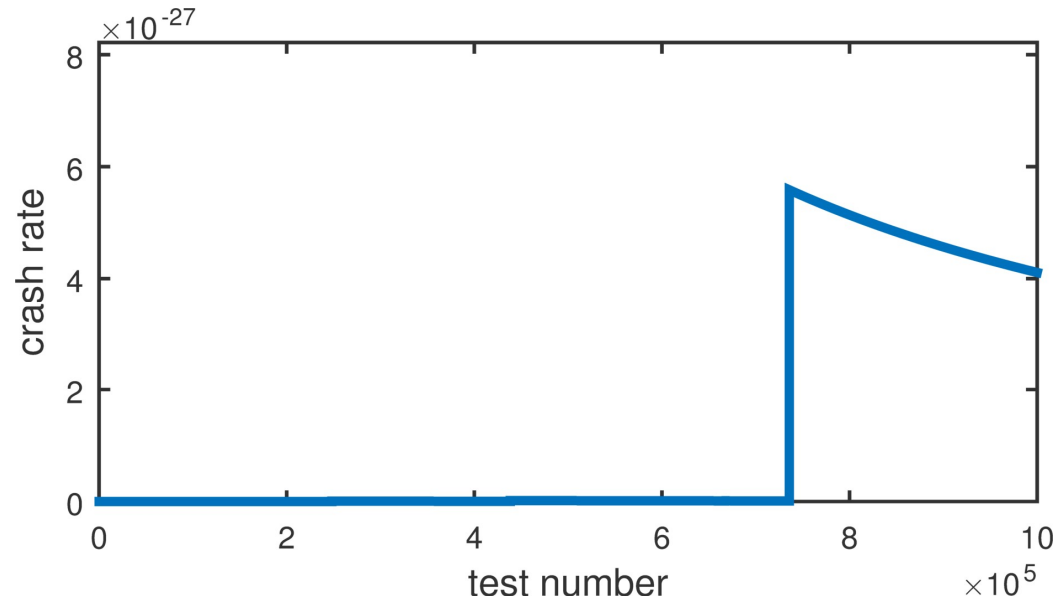

Fig. 8. Estimated crash rate using the uniform distribution

### B. Results of the Proposed Accelerated Evaluation Process

We applied the proposed Accelerated Evaluation approach on three types of events: crash, injury, and conflict events. Both accelerated and naturalistic simulations were conducted. Several examples of the lead HV speed are shown in Fig. 9. In the accelerated simulations, the vehicle tends to accelerate and then decelerate at significant magnitudes. In the naturalistic simulations, the deceleration is much milder. Fig. 10 shows an example in an accelerated simulation that leads to a crash. The crash is caused by frequent acceleration and deceleration, and in this case, a "final blow" is severe braking from a high speed. When the lead HV accelerates, it creates a larger $R_L$. The AV then accelerates to catch up. If the HV harshly brakes at the moment when AV overshoots, it may lead to a crash. This tactic is frequently observed in accelerated simulations, but not in the current Euro-NCAP test protocols or ISO standard tests. In other words, the proposed accelerated evaluation method generates high risk maneuvers, some of which could be considered in future government certification processes.

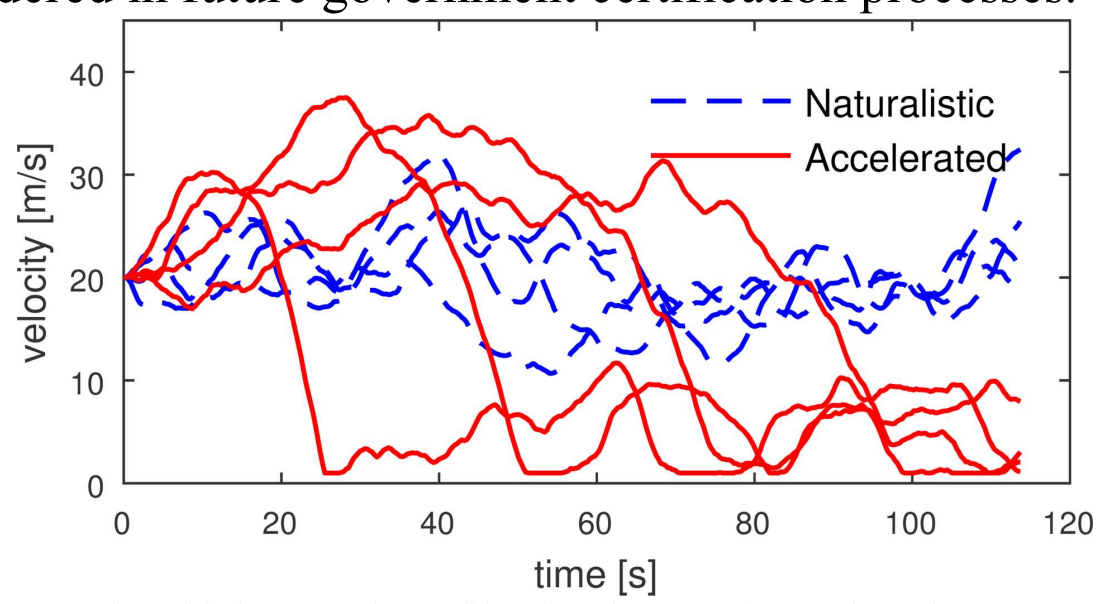

Fig. 9. Lead vehicle speed profiles in the accelerated and non-accelerated (naturalistic driving) simulations

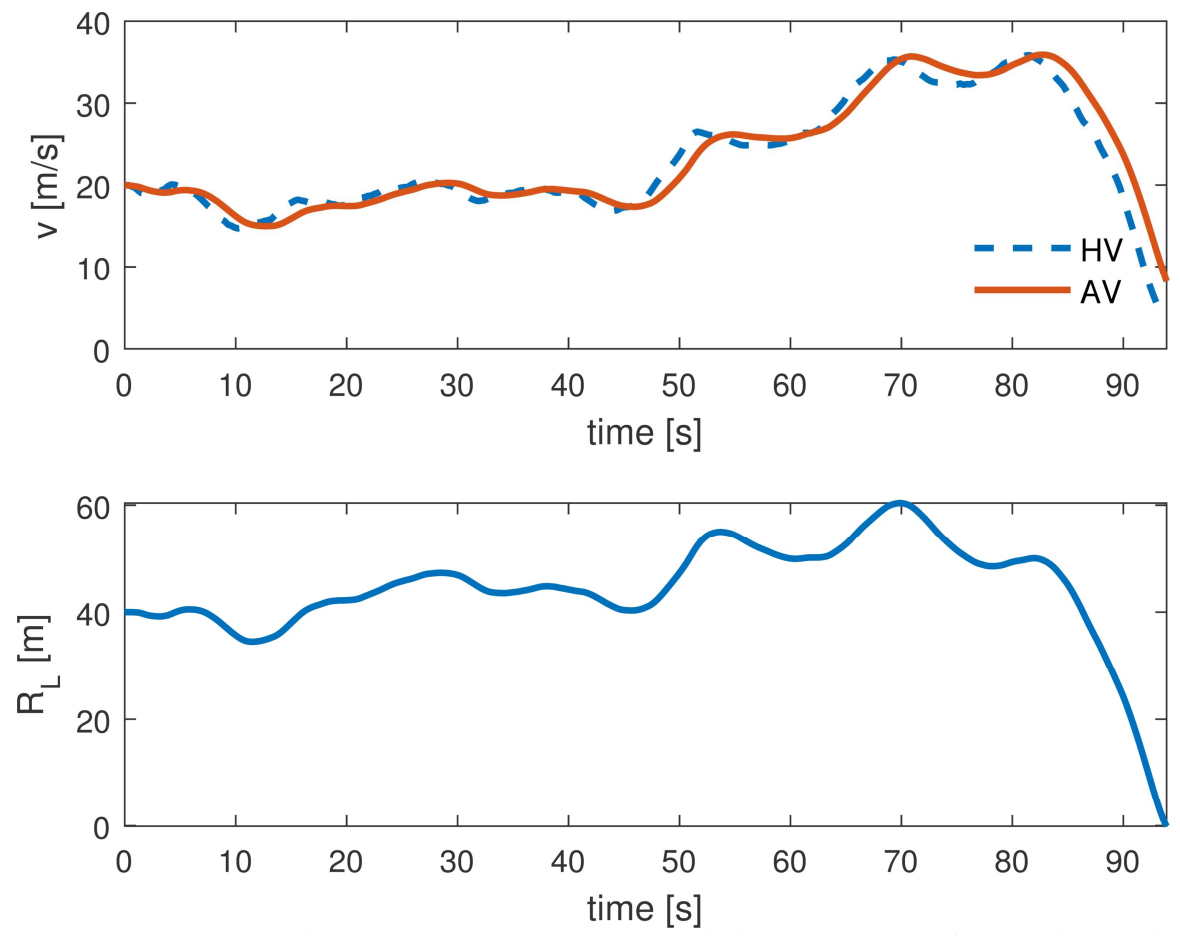

Fig. 10. An example maneuver generated by the accelerated evaluation approach leading to a crash

The crash and conflict events are binary events with indication function defined in (32). When a crash occurs, injury may or may not occur. We model the injury event as a probabilistic function. Here we focus on injuries with the Maximum Abbreviated Injury Score equal to or larger than 2 (MAIS2+), representing moderate-to-fatal injuries. The probability of injury is directly related to the relative velocity

$$\Delta v = -\dot{R}_L(t_{crash}) > 0 \tag{64}$$

The probability of moderate-to-fatal injuries for the AV passengers is estimated by a nonlinear model

$$\mathbb{P}_{\text{inj}}(\Delta v) = \begin{cases} \dfrac{1}{1 + e^{-(\beta_0 + \beta_1 \Delta v + \beta_2)}} & \text{crash} \\ 0 & \textit{no crash} \end{cases} \tag{65}$$

proposed by Kusano and Gabler [61] with $\beta_0 = -6.068$ , $\beta_1 = 0.1$ , and $\beta_2 = -0.6234$ . The injury rate $\mathbb{E}\left(\mathrm{P}_{\text{inj}}(\Delta v)\right)$ is calculated from

$$\mathrm{E}\left(\mathrm{P}_{\text{inj}}(\Delta v)\right) = \hat{\mathrm{E}}_{f^*}\left(\mathrm{P}_{\text{inj}}(\Delta v)\right) \approx \frac{1}{n}\sum_{i=0}^{N_{acc}} \mathrm{P}_{\text{inj}}(\Delta v(n))L(n) \tag{66}$$

where $L(n)$ is the likelihood in the n[th] simulation.

The estimation of crash, injury, and conflict rate are shown in Fig. 11. The estimated crash rates calculated in the accelerated tests converge to those in the naturalistic driving tests, which demonstrates that the Accelerated Evaluation is unbiased.

Both accelerated and naturalistic simulations were conducted until the crash, injury, conflict rates converged with an 80 % confidence level and $\beta = 0.2$ . The speed of convergences are shown in d), e), and f) in Fig. 11. The accelerated tests converged after $N_{acc}$ simulations, while the naturalistic method needed $N_{nature}$ simulations. TABLE III summarizes the performance of the Accelerated Evaluation in estimating the three metrics of the AV. The accelerated rate $r_{acc}$ is defined as $N_{nature}/N_{acc}$. In the crash and injury cases, the proposed method accelerates the evaluation by five orders of magnitude. In the conflict case, the acceleration rate is lower by three hundred times, likely because the Importance Sampling theory provides a larger accelerated rate when the target events are rarer.

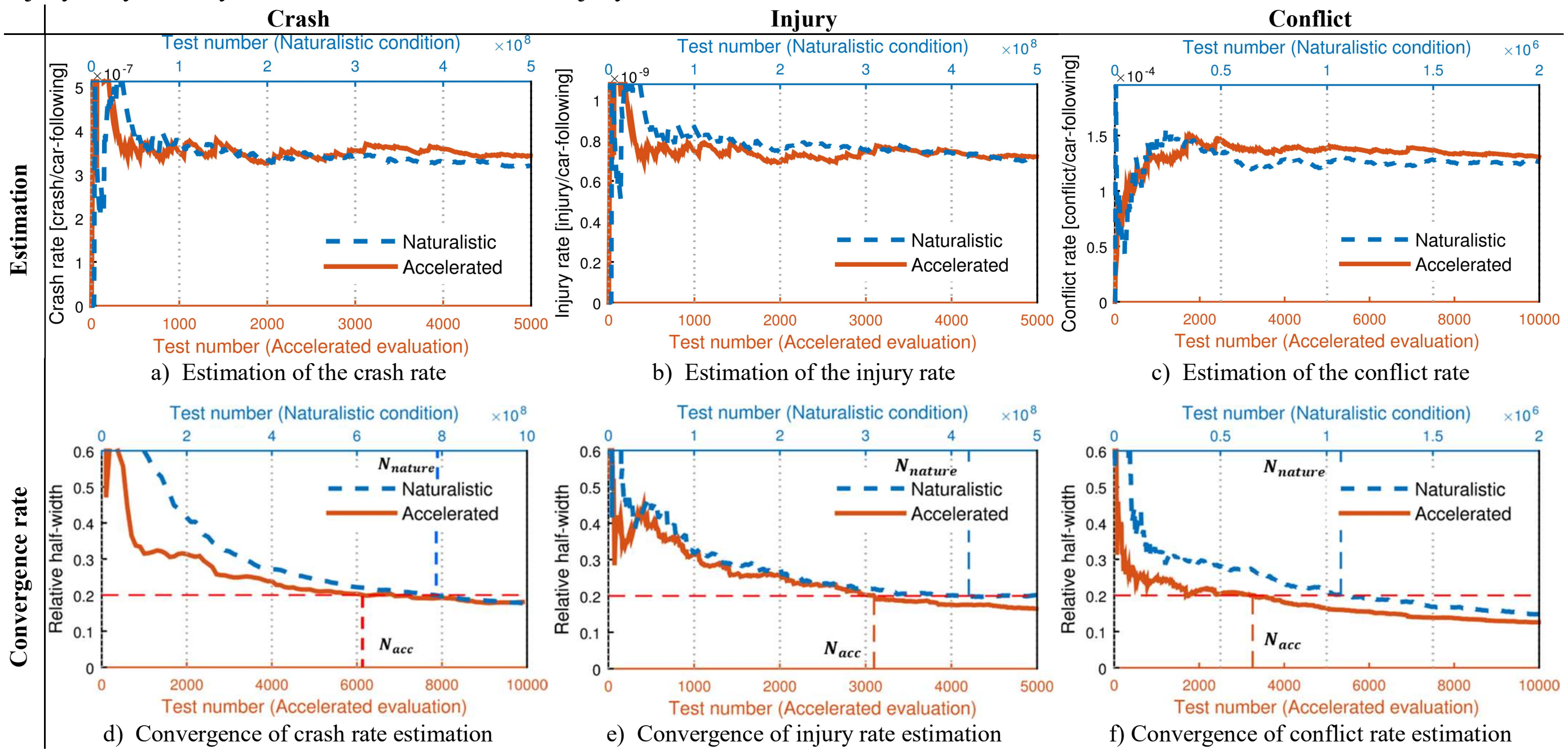

Fig. 11. Estimation and convergence rate of crash, injury, and conflict events

TABLE III
ACCELERATED RATES OF CRASH, CONFLICT AND INJURY EVENTS

| | Crash | Injury | Conflict |
|---|---|---|---|
| $N_{nature}$ | 4.30e8 | 4.20e8 | 1.07e6 |
| $N_{acc}$ | 3.84e3 | 3.10e3 | 3.26e3 |
| $r_{acc} = N_{nature}/N_{acc}$ | 1.12e5 | 1.35e5 | 3.28e2 |

## V. CONCLUSION

An accelerated evaluation method is proposed to assess the safety performance of Automated Vehicles (AV) in the car-following scenario. The statistics of the lead Human Controlled Vehicle (HV) motion is modified so that the interactions between the HV and AV are significantly intensified. A stochastic optimization method is used to minimize the evaluation duration. Importance Sampling theory is then used to estimate the safety performance of AV in the naturalistic driving condition using test results in the accelerated tests.

Simulation results show that the accelerated tests can reduce the evaluation time of crash, injury, or conflict events between 300 and 100,000 times. In other words, simulating for 1,000 miles can expose the AV to challenging scenarios that take 300 thousand to 100 million miles in the real-world to encounter. This technique thus has the potential to dramatically reduce the development and validation time of AVs and can be used by both the government in AV assessment and the auto companies to improve their products.

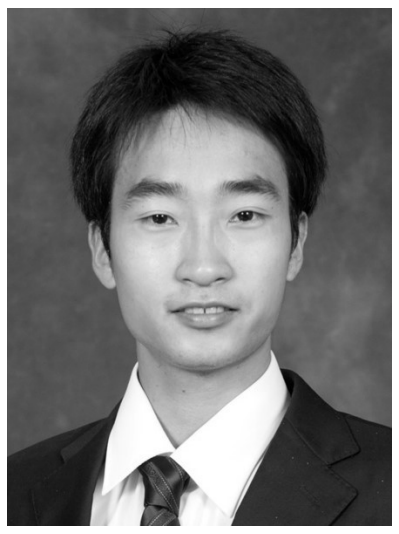

**Ding Zhao** received his Ph.D. degree in 2016 from the University of Michigan, Ann Arbor. He is currently a Research Fellow at the University of Michigan Transportation Research Institute. His research interests include evaluation of connected and automated vehicles, vehicle dynamic control, driver modeling, and big data analysis.

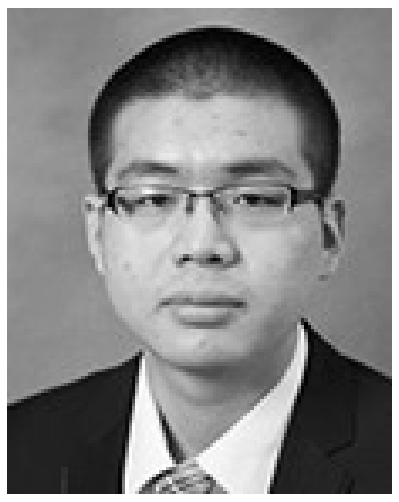

**Xianan Huang** received the B.S. degree in Mechanical Engineering from Shanghai Jiaotong University and Purdue University in 2014. He is currently pursuing the Master's degree in Mechanical Engineering at the University of Michigan, Ann Arbor. His research interests include connected vehicle, hybrid vehicle, and system dynamics and controls.

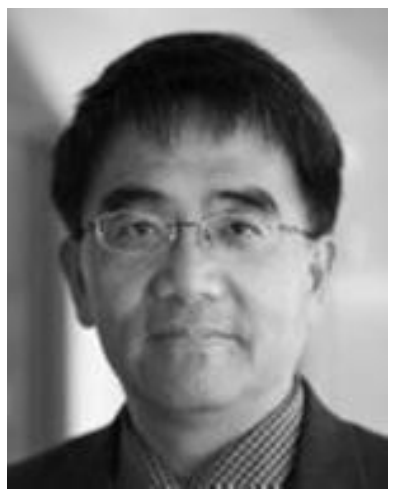

**Huei Peng** received the Ph.D. degree from the University of California, Berkeley, CA, USA, in 1992. He is currently a Professor with the Department of Mechanical Engineering, University of Michigan, Ann Arbor, MI, USA. He has more than 200 technical publications, including 85 in refereed journals and transactions. His research interests include adaptive control and optimal control, with emphasis on their applications to vehicular and transportation systems. His research interests include adaptive control and optimal control, with emphasis on their applications to vehicular and transportation systems.

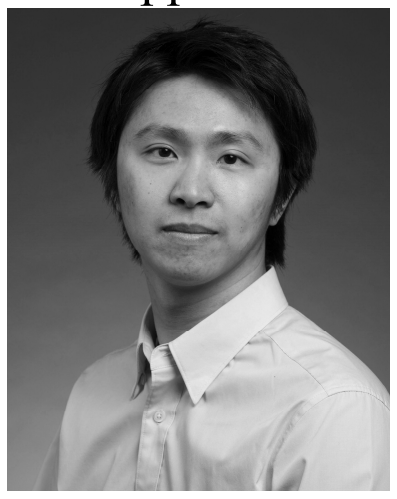

**Henry Lam** received the B.S. degree in actuarial science from the University of Hong Kong in 2005, and Ph.D. degrees in statistics from Harvard University, Cambridge in 2011. Currently, he is an Assistant Professor in the Department of Industrial and Operations Engineering at the University of Michigan, Ann Arbor. His research focuses on stochastic simulation, risk analysis, and simulation optimization. Dr. Lam's works have been funded by National Science Foundation and National Security Agency. He has also received an Honorable Mention Prize in the Institute for Operations Research and Management Sciences (INFORMS) George Nicholson Best Student Paper Award, and Finalist in INFORMS Junior Faculty Interest Group Best Paper Competition.

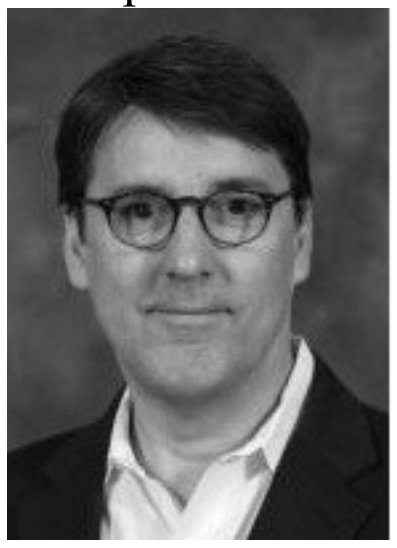

**Dave LeBlanc** received a Ph.D. in aerospace engineering from the University of Michigan, and master's and bachelor's degrees in mechanical engineering from Purdue University. Dr. David J. LeBlanc is currently an associate research scientist, has been at UMTRI since 1999. Dr. LeBlanc's work focuses on the automatic and human control of motor vehicles, particularly the design and evaluation of driver assistance systems.